# Resilience in urban networked infrastructure: the case of Water Distribution Systems

Antonio Candelieri[1], Ilaria Giordani[3], Andrea Ponti[2], Riccardo Perego[2] and Francesco Archetti[2]

[1] Department of Economics, Management and Statistics,
University of Milano-Bicocca, 20126, Milan, Italy
[2] Department of Computer Science, Systems and Communication,
University of Milano-Bicocca, 20126, Milan, Italy
[3] Oaks srl. Milano
`antonio.candelieri@unimib.it`

**Abstract.** Resilience is meant as the capability of a networked infrastructure to provide its service even if some components fail: in this paper we focus on how resilience depends both on net-wide measures of connectivity and the role of a single component. This paper has two objectives: first to show how a set of global measures can be obtained using techniques from network theory, in particular how the spectral analysis of the adjacency and Laplacian matrices and a similarity measure based on Jensen-Shannon divergence allows us to obtain a characterization of global connectivity which is both mathematically sound and operational. Second, how a clustering method in the subspace spanned by the l smallest eigenvectors of the Laplacian matrix allows us to identify the edges of the network whose failure breaks down the network.
Even if most of the analysis can be applied to a generic networked infrastructure, specific references will be made to Water Distribution Networks (WDN).

## 1 Introduction

Networked infrastructures, as water, energy and transport, have developed similar functional and structural features in their evolution over time: spatial, but also financial, constraints have significantly restricted their connectivity, robustness and their capability to deliver their service with failed or damaged components, in short their resilience. These features have also generated systemic risk and cascading effects exacerbated by the complexity of the infrastructure with up to tens of thousands of components (pipes, valves, pumping stations, tanks and consumption points). Resilience of a Water Distribution Network (WDN) is about delivering services regardless of disruptive events that may occur. *Resilience*, *robustness*, *reliability* and *vulnerability* are terms strictly linked and often confusingly used.

We propose a framework based on network theory to address structural analysis of any WDN: the growing awareness of the interplay between *global* (system-wide) and *local* (individual component) resilience have spawned a line of research directly aimed at resilience in WDN (Yazdani and Jeffrey, 2012), (Soldi et al. 2015).

Complex networks are instances of real-world graphs. They include examples such as the Internet, social networks, supply networks, metabolic networks, and critical infrastructures, among other engineered systems. Important characterizations are: *small-world* networks (Backstrom et al., 2012), *scale-free* networks (Barabási et al., 2009) and *planarity*, which typically characterizes road networks, water distribution networks, energy grids and general networked flow systems. Another important characterization, relevant for resilience, is the *community structure* (Girvan et al., 2002), meaning that connections are dense among nodes within the same subset (i.e., "community") and sparse among nodes between different subsets.

## 1.1 Related Work

Graph theoretic approaches have been proposed in the literature to address the issue of resilience in WDN both in terms of connectivity and service levels (Diao et al., 2016), (Shuang et al., 2014), (Archetti et al., 2015), (Candelieri et al., 2017), (Gutiérrez-Pérez et al., 2013), (Herrera et al., 2016), (Di Nardo et al., 2018), (Diao et al. 2020). In this paper we focus on *connectivity*, whose analysis in network models offers important cues to the design and management of the network even without the need of running a hydraulic simulation model. *Spectral analysis* of networks offers a mathematically principled approach yielding both local and global structural information at a computational cost of $O(n^3)$, with $n$ the number of nodes of the network, due to the computation of eigenvalue and eigenvectors. An innovative perspective on the structural characterization of networks is offered in (Schieber et al., 2017), based not only on averages but on distributional properties analyzed by information theoretic models. A related approach to the evaluation of resilience, drawn from physics, is provided by *percolation analysis* which evaluates the impact of removing nodes/links from the network in terms of how the average length of the shortest paths connecting pairs of nodes increases, to the point of bringing to a disconnected network. Monte Carlo (MC) methods are key for percolation in complex networks (Chen, 2017) in which several random global disruption scenarios are analyzed (Torres et al., 2017). A simulation-based model percolation is flexible and can handle several kinds of network failures ranging from a single node to a scenario in which a critical fraction of the network components has failed.

4## 1.2 Our Contributions

The main contribution of this paper is a global approach to the characterization of resilience in urban water distribution network (WDN). Specific contribution is articulated in 3 points:

- To show how a set of global measures can be obtained using techniques from network theory, in particular how the spectral analysis of the adjacency and Laplacian matrices allows us to obtain a characterization of global connectivity which is both mathematically sound and operational.
- To show that considering the analysis of the node-node distance distribution and the *Network Node Dispersion* (NND) (Schieber et al., 2017) can yield additional insights on network structural characterization.
- To show how a graph-clustering method, in the subspace spanned by the $l$ smallest eigen-vectors of the Laplacian matrix, allows us to identify the edges whose failure breaks the network into unconnected components.

The proposed approach has been evaluated on both benchmark and real world WDNs, considering breakages on pipes as relevant disruptive events. The structure of the paper is as follows: section 2 gives background notions on graph models and network analysis; section 3 describes the measures and tools provided by spectral analysis. Section 4 describes the different WDNs used in this study and the relevant results; finally, in section 5 some conclusions are provided.

## 2 Mathematical Background

## 2.1 Graph Theory

Let denote a graph with $G = (V, E)$, where $V$ is the set of nodes and $E$ is the set of edges. Each edge of $G$ is represented by a pair of nodes $(i, j)$ with $i \neq j$, and $i, j \in V$ and with $n = |V|$ and $m = |E|$. If $(i, j) \in E$, $i$ and $j$ are called *adjacent* nodes. A graph $G$ is *undirected* if $(i, j)$ and $(j, i)$ represent the same edge. A graph $G$ is *simple* if no self-loops are admitted (edges starting from a node and ending on the same node) and only one edge can exist between each pair of nodes $(i, j)$, with $i \neq j$. The *adjacency* relationship between the nodes of $G$ can be represented through a non-negative $n \times n$ matrix $A$ (i.e., the adjacency matrix of $G$). The entry $a_{i,j}$ of the adjacency matrix $A$ is 1 if $i$ and $j$ are adjacent nodes (i.e., $(i, j) \in E$), and 0 otherwise. Furthermore, $a_{ij} = a_{ji}$ if $G$ is undirected and $a_{ij}$ (entries on the diagonal) are 0 if $G$ is simple.

Let denote with $k_i$ the degree of the node $i$, that is the number of edges having $i$ as one of the two nodes on the edge $k_i = \sum_{j=1}^{n} a_{i,j}$. Anyone of the edges having $i$ as one of its nodes is called incident on $i$.

When $G$ is *directed*, meaning that the order of the two nodes of an edge is relevant for its definition, the $k_i$ can be split into *out-degree* (number of edges having $i$ as first node) and *in-degree* (number of edges having $i$ as second node).

A *path* in a graph is a sequence of nodes connected by edges the length of the path is the number of edges. A connected component is a maximal subgraph when all nodes can be reached from every other.

*The shortest path* between $i$ and $j$ is the one related to the smallest number of arcs from $i$ to $j$, which is usually named distance $d(i,j)$. The largest distance among each possible pair of nodes in $G$ is named diameter.

A subgraph $G' = (V', E')$ of $G$ is a graph such that $V' \subseteq V$ and $E' \subseteq E$; a connected component of $G$ is a maximal if is the largest possible graph for which you could not find another node in the graph that could be added to the graph with all the nodes be still connected..

A weight $w_{ij} \geq 0$ can also be associated with every edge $(i,j) \in E$; in this case the graph $G$ is called *weighted* and the (weighted) adjacency matrix is a $n \times n$ matrix $W$ having $w_{ii} = 0$, if $G$ is simple, $w_{ij} \geq 0$ and $w_{ij} = w_{ji}$ for each $i \neq j$ if $G$ is undirected. In the case of weighted graphs, the previous definitions, related to degree, path and diameter, can be modified to consider weights on the edges rather than their number. In particular, degree of the node $i$ is the sum of the weights of the edges incident on $i$ (out-degree is the sum of the weights of the edges starting from $i$, while in-degree is the sum of the weights of the edges ending to $i$); shortest path between $i$ and $j$ is the list of adjacent nodes from $i$ to $j$ with minimal sum of the weights on the correspondent connecting edges; the diameter is the longed shortest path computed.

## 2.2 Network Analysis: the basic measures

The number of edges $m = 1/2 \sum_{i=1}^{n} k_i = \frac{1}{2} \sum_{i,j=1}^{n} a_{ij}$.

If $c$ is the mean vertex degree, $c = \frac{1}{n} \sum_{i=1}^{n} k_i$ we get $c = \frac{2m}{n}$.

Since the max possible number of edges in $G$ is $\binom{n}{2} = \frac{n(n-1)}{2}$ we can compute the density of the network as the fraction of edges which are present in the specific graph:

$$q = \frac{m}{\binom{n}{2}} = \frac{2m}{n(n-1)} = \frac{c}{n-1}$$

the density is in the range (0,1).

A pair of nodes is usually connected by many paths which typically share some nodes or edges. If they share no edges, they are called *edge independent*. No shared nodes imply *vertex independence*. The number of independent paths between 2 nodes is called *connectivity* of the 2 nodes. A *cut-set*, specifically a vertex cut-set, is a set of nodes whose removal disconnects $i$ and $j$. A *minimum cut-set* is the smallest cut-set.

An important concept is the Laplacian matrix of a network $L = D - A$, where $A$ is the adjacency matrix and $D$ is a $n \times n$ diagonal matrix with $d_{ii} = k_i$ (Brouwer et al., 2011). The eigenvalues of $L$ are of paramount importance in assessing the connectivity. We number them as $0 = \lambda_1 \leq \lambda_2 \leq \cdots \leq \lambda_n$, so they are not negative. Note that $L$ is singular. If we have $h$ different components of size $n_1, n_2, \ldots, n_h$, then $L$ is block diagonal and the multiplicity of the zero-eigenvalue is exactly equal to the number of components, which in turn implies that $\lambda_2$ is non zero if and only if the network is connected; $\lambda_2$ is also called *algebraic connectivity*.

Also important in the analysis of resilience are *centrality measures*, which address the issue of the relative importance of nodes/edges. The most widely used measures are:

*Eigenvector centrality* of the vertex $i$, that is $X_i = \varphi_1^{-1} \sum_{j=1, j \neq i}^{n} a_{ij} x_j$, where $\varphi_1$ is the largest eigenvalue of the adjacency matrix, $A$, (aka *spectral radius*). The eigenvector centrality can be large either because the vertex has many neighbors or because has important neighbors. *Katz centrality* and *Page Rank* algorithm are just parametrized version of eigenvector centrality.

*Closeness centrality* measures the mean distance from one vertex to the others. Let $d_{ij}$ be the length of a shortest path from $i$ to $j$, that is the number of edges along that path. The closeness centrality is: $C = \frac{n}{\sum d_{ij}}$.

*Betweenness centrality*: let be $\eta_{st}^i = 1$ if vertex $i$ lies on the shortest path from $s$ to $t$ and 0 otherwise. Then, betweenness centrality is given by $b_i = \frac{1}{n^2} \sum_{s,t=1}^{n} \eta_{st}^i$.

We can similarly define an *edge betweenness* that counts the number of shorter paths that run along the edges. Upon these indices we can build a first characterization of resilience removing the vertex/edge with the highest centrality score until the network splits.

As a basic measure of connectivity, the *average degree* can provide an immediate information about the organization of the network. This measure is also linked to the *link-per-node ratio (e)*, that is computed as the number of edges of a graph with respect to the number of its nodes.

*Central point dominance* $c_b'$, based on betweenness centrality is a measure for characterizing the organization of a network according to its path-related connectivity; $c_b' = \frac{1}{n-1} \sum_{i=1,\ldots,n} (b_{max} - b_i)$
where $(b_i)$ is the betweenness centrality of the node $i$ and $b_{max}$ is the maximum value of betweenness centrality over all the n nodes of the network.

The evaluation of network resilience requires to extend the analysis other structural features: the *clustering coefficient (CC)* is used to characterize the resilience of a network according to loops of length three and is computed as the number of triangles with respect to the overall number of possible connected triples, where a triple consists of three nodes connected at least by two edges while a triangle consists of three nodes connected exactly by three edges:

$$CC = \frac{3N_{triangles}}{N_{triples}}$$

In this paper, the open-source software Cytoscape (http://www.cytoscape.org/) has been adopted as the basic tool of the analytical framework proposed and its plug-in-ClusterMaker2 (Morris et al., 2011).

## 2.3 Dissimilarity analysis &Network Nodes Dispersion (NND)

The measures introduced in Section 2.2 are based on distances and their average values. Another analysis can be performed also in terms of distributions. This kind of analysis has been inspired by the paper (Schieber et al., 2017) which is based on the vertex-vertex distance distribution. The first step is to consider a measure of the graph heterogeneity through connectivity distances. The shortest path distances between all nodes are arranged in the distance matrix $T = [t_{i,j}], i,j = 1, \dots, n$.

The maximum entry of row $i$, $\max_{j=1,\dots,n} t_{i,j}$ is known as the *eccentricity* of node $i$.

The maximum eccentricity among the nodes $\max_{i,j} t_{i,j}$ is equal to the diameter of the network.

For each row $i$ we compute $p_i(j)$ as the fraction of nodes which are connected to $i$ at a distance $j$ and associate to node $i$ the probability distribution $P_i$ of the r.v. $p_i(j)$.

The Network Node Dispersion (NND) is given by

$$NND(G) = \frac{J(P_1, \dots, P_n)}{\log(d+1)}$$

The *Jensen-Shannon* divergence of the probability distributions $P_i, i = 1, \dots, n$ is given by

$$\sum_{i,j} p_i(j) \log\left(\frac{p_i(j)}{\mu_j}\right)$$

where $\mu_j = \sum_{j=1}^{n} \frac{p(j)}{n}$ and is normalized by $\log(d+1)$ where $d$ is the diameter of the network.

Considering the distance distribution over the whole graph we obtain the average node distance distribution $P(G)$ with average $\mu_G$.

This enables to compute a measure of similarity with another graph $G'$ through the Jensen-Shannon divergence $J(P(G), P(G'))$.

Then we measure the distance between $G$ and $G'$ by

$$D(G, G') = w_1 \sqrt{\frac{J(P(G),P(G'))}{\log 2}} + w_2 \left|\sqrt{NND(G)} - \sqrt{NND(G')}\right|$$

with $w_1 + w_2 = 1$.

This distance is different than the one used in Scheiter (2016) which has a third term which takes into account the centrality measures of each node and its connectivity span.

We use instead the set of centrality related measures introduced in 2.2.

## 3 Spectral Clustering

Given two sets of nodes $C_1$ and $C_2$, an $n$-dimensional vector $z$ i.e., $n$ is the number of nodes in the graph) is used to represent the association of each node to cluster $C_1$ or $C_2$

$$z_i = \begin{cases} +1 \text{ if } i \in C_1 \\ -1 \text{ if } i \in C_2 \end{cases}$$

The graph clustering problem can be formulated as the minimization of the following function $f(z)$:

$$f(z) = \sum_{x_i, x_j \in V} L_{ij}(z_i - z_j)^2 = z^T L z$$

where $L_{ij}$ are the entries of the Laplacian matrix.

The important feature of spectral clustering methods is that the produce a set of balanced clusters. An elegant solution, conceptually simple but computationally inefficient, to the problem was proposed in (Fiedler, 1973) which identified the 2nd smallest eigenvector of the Laplacian matrix (usually known as Fiedler vector) as the vector $p$ which provides the optimal bi-partitioning of the graph. Early applications of this result have permitted to implement recursive bi-partitioning spectral clustering approaches (Hagen and Kahng, 1992) to perform partitioning in $K > 2$ groups. However, this approach requires the computation of matrices, eigenvalues and eigenvectors, for each sub-graph until the desired number of clusters is reached. More effective computational schemes are analyzed in (Luxburg, 2007) and use a data representation in the lower dimensional space spanned by the most relevant eigenvectors. Our approach in this paper consists in ranking in descending order the eigenvalues $\varphi_1 \geq \varphi_2 \geq \cdots \geq \varphi_n$ of the adjacency matrix. If the user sets the desired number of clusters as k, k-means clustering is performed on the resulting dataset having *n* rows (nodes of the graph) and *k* columns (eigenvectors corresponding to the *k largest* eigenvalues of A). If a suitable value of k is not known the implementation in Cytoscape ClusterMaker2 computes the ratio $\varphi_i / \varphi_{i+1}$, $i = 1, \ldots, n-1$ and picks as $k$ the smallest integer $i$ such the ratio than 1+ε (in the computation reported in Section 4, $\varepsilon = 1.02$).

## 4  Experimental Setting

### 4.1  The Network Models

In this section 4 WDNs are analyzed. The first WDN is a benchmark model often used in different studies on WDN management, namely "Anytown"1. The associated graph consists of 22 nodes and 43 links.

Marnate is a small town in Northern Italy, with an associated graph consisting of 384 nodes and 469 edges.

Neptun is the WDN of the Romanian city of Timisoara, with an associated graph of 333 nodes and 339 edges, analyzed in the European project Icewater.

Abbiategrasso refers to a pressure management zone in Milan (namely, Abbiategrasso) with an associated graph consisting of 1212 nodes and 1385, analyzed in the European project Icewater.

In analyzing WDNs one must consider that most of the end-users are supplied by single connections. To avoid a bias in the analysis, a preliminary preprocessing can be performed by cutting the final connections, that are usually the links between the consumption meters of each building and the main distribution pipes.

### 4.2  Computational results

The characteristic path length is the average number of edges along the shortest path for every possible pair of nodes $(i,j)$.

**Table 1.** Structural Analysis of four WDNs

| Measure | Anytown | Marnate | Neptun | Abbiategrasso |
|---|---|---|---|---|
| **Density (q)** | 0.186 | 0.006 | 0.005 | 0.001 |
| **Link-per-node ratio (e)** | 1.954 | 2.443 | 0.992 | 1.156 |
| **Central point dominance (cb')** | 0.230 | 0.189 | 0.476 | 0.303 |
| **Clustering coefficient (CC)** | 0.303 | 0.007 | 0.000 | 0.004 |
| **Diameter** | 7 | 35 | 82 | 83 |
| **Characteristic Path Length** | 2.761 | 21.696 | 30.226 | 31.233 |

Anytown looks rather more like a "No-town" network, with structural properties far from those of the real WDNs. The three real-world WDNs analyzed are very sparse (with density $q$ lower or equal to 0.006) with respect to Anytown (density q around 0.2). The central point dominance $c_b'$, instead, is quite similar among all the four WDNs taken into account. The clustering coefficient $CC$, diameter d and characteristic path length are quite similar among the three real WDNs and different from those computed on Anytown: the three real-world WDNs are effectively planar and "almost" regular.

---

[1]  http://emps.exeter.ac.uk/engineering/research/cws/resources/benchmarks/expansion/anytown.html

Table 2. Spectral Analysis of four WDNs

| Measure | Anytown | Marnate | Neptun | Abbiategrasso |
|---|---|---|---|---|
| **Spectral Gap** | 1.5149 | 0.0838 | 0.0149 | 0.2132 |
| **Algebraic Connectivity** | 0.1708 | 0.0046 | 0.0009 | 0.0002 |

The spectral gap is the difference between the two largest eigenvalues of the Laplacian matrix. The algebraic connectivity is the value of $\lambda_2$.

The spectral analysis shows that the 3 real world WDN have relatively similar values and again quite different from Anytown.

Table 3. Dissimilarity of four WDN

| Measure | Anytown | Marnate | Neptun | Abbiategrasso |
|---|---|---|---|---|
| **Anytown** | 0.000 | 0.670 | 0.700 | 0.770 |
| **Marnate** | 0.670 | 0.000 | 0.335 | 0.681 |
| **Neptun** | 0.700 | 0.335 | 0.000 | 0.617 |
| **Abbiategrasso** | 0.770 | 0.681 | 0.617 | 0.000 |

Anytown is different from the others and the difference is captured quite naturally. Marnate and Neptun are quite similar and different from Abbiategrasso. Actually, Neptun and Marnate have grown out of autonomous urban water distribution networks, constrained in their development by technological and physical constraints. Abbiategrasso is a subnetwork – specifically a Pressure Management Zone – carved out of the whole water distribution network of Milano for administrative reasons and management strategies. This structural difference is captured by the dissimilarity measure.

### 4.3 Clustering

Graph clustering approaches, such as Spectral Clustering, can be used to identify the specific links (pipelines) whose removal may induce a disconnection of the network in two or more sub-networks. In this paper, Spectral Clustering has been performed (through Cytoscape's Cluster plug-in named ClusterMaker2) to identify sub-networks connected by a limited (minimal) number of links, that are pipelines whose breakage implies the disconnection of some WDN portion. In the following figures these pipelines are highlighted; it is important to note that breakages must occur, at the same time, on all the different red edges to imply a hydraulic disconnection. Breakages affecting only one pipe may imply a reduction in the supply service or generate a "stress" condition on the hydraulic infrastructure. A software simulation of the damaged network may be used to evaluate the induced scenario.

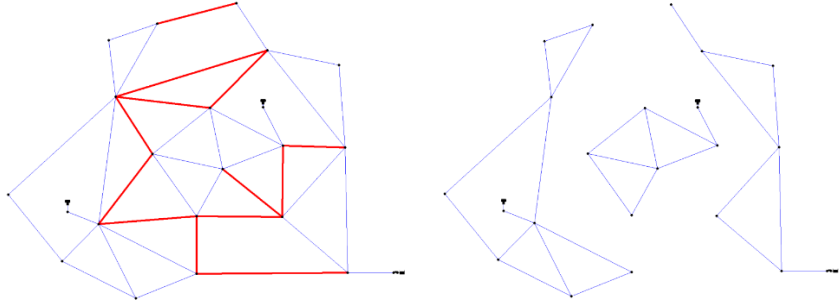

**Fig. 1**. Anytown (k=3): Critical edges (red) whose removal generates a disconnection and (right) resulting disconnected components.

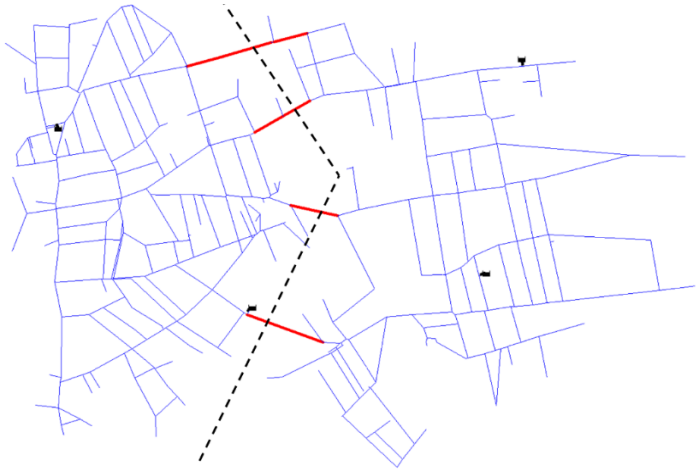

**Fig. 2**. Marnate WDN (k=3): Critical edges (red) whose removal generates a disconnection.

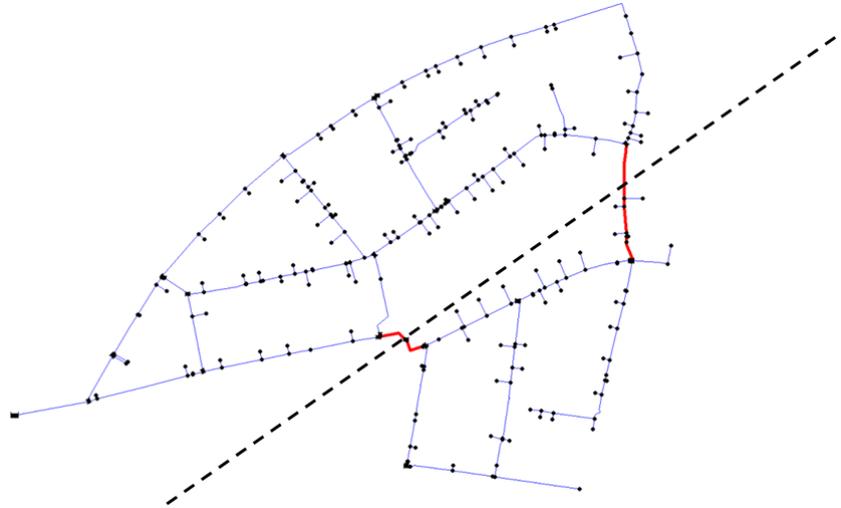

**Fig. 3.** Neptun WDN (k=2): Critical edges (red) whose removal generates a disconnection.

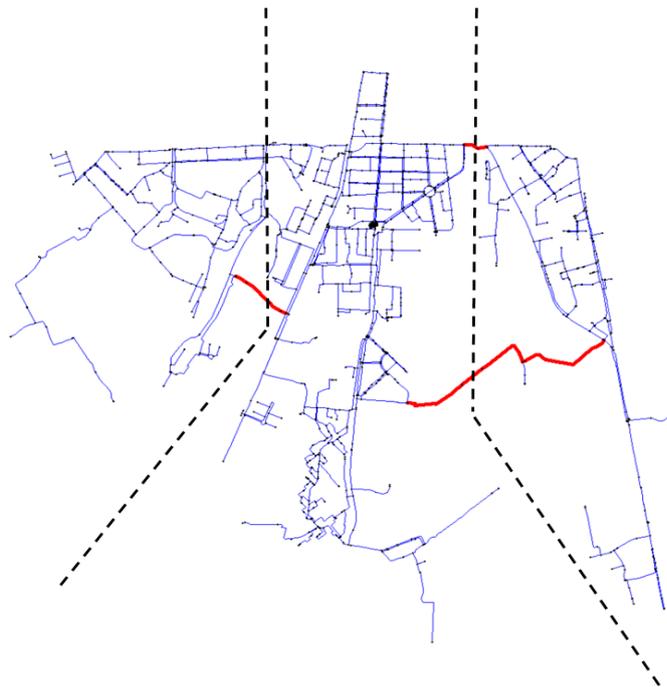

**Fig. 4.** Abbiategrasso WDN (k=3): Critical edges (red) whose removal generates a disconnection.

According to results of Spectral Clustering, the disconnection in two sub-networks is reported for the Marnate WDN, while the disconnection in three sub-networks is depicted for both Bresso-Cormano-Cusano and Abbiategrasso WDNs. More in detail, respect to Anytown, Spectral Clustering is not able to provide a bi-partitioning of the WDN in a reasonable time, mainly due to the high connectivity of the water network. Moreover, the disconnection in three different sub-networks may occur only by the simultaneous breakage of many pipelines.

## 5  Conclusion

In this paper the use of network analysis for the evaluation of resilience in a urban networked infrastructure has been proposed.

The application to three WDNs from two different projects and one of benchmark has permitted to define the main measures and their characteristic values for real world WDNs, taking into account also previous results reported in the literature.

While general measures have been used in order to evaluate and compare connectivity and resilience of the WDNs considered, the application of spectral clustering has permitted to identify the most critical hydraulic pipelines whose breakage imply structural disconnection and consequent failure of the distribution service (vulnerability).

A further layer of analysis that can be added consists in joining the network analysis, in the abstract graph setting, and hydraulic simulation, provided for instance by EPANET. The set of resilience indices based on network analysis, and adopted in this paper, continues to measure how the failure of a single component impacts the connectivity while the simulation of the damaged network provides a measure about how a damaged component impacts the service level still offered by the WDN.

## Acknowledgements

This study has been partially supported by the Italian project "PERFORM-WATER 2030" – programma POR (Programma Operativo Regionale) FESR (Fondo Europeo di Sviluppo Regionale) 2014–2020, innovation call "Accordi per la Ricerca e l'Innovazione" ("Agreements for Research and Innovation") of Regione Lombardia, (DGR N. 5245/2016 - AZIONE I.1.B.1.3 – ASSE I POR FESR 2014–2020) – CUP E46D17000120009.

We greatly acknowledge the DEMS Data Science Lab for supporting this work by providing computational resources (DEMS – Department of Economics, Management and Statistics).